# Revealing the nature of defects in quasi free standing monolayer graphene on SiC(0001) by means of Density Functional Theory

Tommaso Cavallucci,[1] Yuya Murata,[1] Makoto Takamura,[2] Hiroki Hibino,[2,†] Stefan Heun,[1] Valentina Tozzini[1,*]

[1] NEST, Istituto Nanoscienze-CNR and Scuola Normale Superiore, Piazza San Silvestro 12, 56127 Pisa, Italy

[2] NTT Basic Research Laboratories, 3-1 Morinosato Wakamiya, Atsugi, Kanagawa 243-0198, Japan

**ABSTRACT:** Quasi free standing monolayer graphene (QFMLG) grown on SiC by selective Si evaporation from the Si-rich SiC(0001) face and H intercalation displays irregularities in STM and AFM analysis, appearing as localized features, which we previously identified as vacancies in the H layer coverage [Y Murata, et al. Nano Res, in press, DOI: 10.1007/s12274-017-1697-x]. The size, shape, brightness, location, and concentration of these features, however, are variable, depending on the hydrogenation conditions. In order to shed light on the nature of these features, in this work we perform a systematic Density Functional Theory study on the structural and electronic properties of QFMLG with defects in the H coverage arranged in different configurations including up to 13 vacant H atoms, and show that these generate localized electronic states with specific electronic structure. Based on the comparison of simulated and measured STM images we are able to associate different vacancies of large size (7-13 missing H) to the 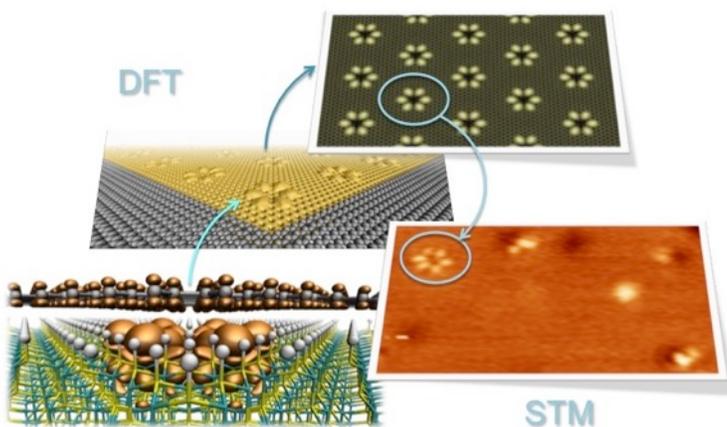 different observed features. The presence of large vacancies is in agreement with the tendency of single H vacancies to aggregate, as demonstrated here by DFT results. This gives some hints into the hydrogenation process. Our work unravels the structural diversity of defects of H coverage in QFMLG and provides operative ways to interpret the variety in the STM images. The energy of the localized states generated by these vacancies is tunable by means of their size and shape, suggesting applications in nano- and opto-electronics.

## 1. Introduction

Thermal decomposition of silicon carbide (SiC) is a widely used technique to produce supported graphene[1]. Upon selective evaporation of Si from the SiC(0001) face, excess carbon produces in the first instance a hexagonal carbon layer, called buffer layer,[2] covalently bound to the substrate, which retains a high amount of sp$^3$ hybridized sites. To obtain real fully sp$^2$ hybridized graphene, Si evaporation can be continued, which leads to the formation of a second buffer layer under the first one and its subsequent detachment, resulting in a graphene monolayer. This system was widely characterized, experimentally,[3,4] and theoretically[1,5,6,7] revealing interesting electronic and structural properties useful for advanced applications[8,9]. Alternatively, the buffer layer can be detached from the SiC substrate by H-intercalation between the buffer layer and the substrate[10], obtaining the so-called Quasi Free Standing Monolayer Graphene (QFMLG)[11]. The morphological properties of this system are less well defined, because they depend on the environmental conditions during intercalation. Specifically, Scanning Tunneling Microscopy (STM) and Spectroscopy (STS) studies reveal the presence of features extending over a few C sites superimposed on the regular graphene atomic structure[12]. In our previous work[13] we showed that these features formed in favorable H-intercalation conditions (high temperature and pressure) have relatively small and voltage dependent contrast (i.e. they appear darker or brighter with respect to the flat graphene areas depending on the chosen bias voltage). By means of a comparative study based on STM and STS experiments and Density Functional Theory (DFT) calculations, we associated them with defects in the H coverage compatible with vacancies with 3 and 4 missing H atoms. These two types of vacancies also display different peak energies in the STS spectra (at +0.9 and +1.2 eV with respect to the Dirac point, respectively). In Ref. [12] it was reported that in given conditions these vacancies tend to align and arrange at a minimum distance of ~1.8±0.1 nm between each other, approximately locating on the 6×6 sites of SiC.

Though the basic nature of the observed features has been demonstrated in Ref. [13], different aspects remain to be clarified. The energetics of formation, surface concentration and distribution of vacancies appear dependent on the conditions in which QFMLG is formed: besides the two main types of vacancies previously described, other features of different size, shape and concentration are observed by STM upon subsequent annealing of the sample or intercalation in different conditions (unpublished data, reported and described hereafter), whose electronic properties were not explored. These circumstances call for a systematic study of the structural and electronic properties of H coverage vacancies of different size, shape and location. To our knowledge, only a couple of DFT studies were performed on QFMLG with full H coverage[14,15], therefore with no indication about the H vacancies behavior. Besides our previous work[13], a DFT study about dangling bonds interacting with graphene was performed only on a different system – i.e. graphene on the C-face of SiC[16].

In this work, we perform a DFT study of QFMLG with different types of H coverage defects, sizing from 1 to 13 missing atoms, arranged in different geometries. Structural, energetic and electronic properties are evaluated and discussed in light of the available STM data from Ref. [13] and of some new data reported here for the first time.

## 2. Materials, Methods and Models

### 2.1 Theoretical models and methods

*The simulation supercells and QFMLG model systems*

A supercell compatible with the symmetry and the lattice parameter of graphene is the 13×13 of graphene superimposed on the 6√3×6√3 R30 of SiC. This was previously used for DFT studies and is considered the "standard" supercell for SiC[5] supported graphene. Indeed, this model system is shown to well reproduce the rippling and other structural characteristics of the graphene buffer and monolayer. It is reported in Fig 1a, left part (exact lattice parameters and other details included in the Fig) and called hereafter the L (large) model system.

A simpler model can however be build, which retains the main observed characteristics, based on rotated supercells of SiC and graphene, namely the 7×7 R21.787 graphene on √31×√31 R8.95 of SiC, represented in Fig 1a (right part, hereafter named S (small) model system). This model system displays only a 0.7 deg of relative rotation of graphene with respect to SiC. The unit cell size is 17.11Å, only ~1 Å smaller than the 6×6 unit cell. When a single defect is located in this super-cell, the distance between periodic copies is compatible with the experimentally observed distance between vacancies. Overall, these observations make the S model an attractive alternative to the L model, and it was in fact used in our previous work[13].

A great advantage of S model is that its size is ~1/3 with respect to L model, corresponding roughly to one order of magnitude gain in computational speed. This allowed us to make extensive and systematic studies of different types of vacancies. Besides the small relative rotation of graphene with respect to SiC, other differences between S and L models are: i) graphene is slightly more contracted in the S cell (about 0.4%, negligible contraction in the L cell); ii) the S model system can host 0 or 1 vacancies per cell, therefore it can reproduce only the system with "maximum" or null vacancy concentration, and the distance between vacancies is fixed by the cell periodicity i.e. ~1.7nm; conversely, the L supercell can host 0, 1, 2 or 3 vacancies, allowing to simulate also lower vacancy concentrations and larger vacancy distances. In this work the S model is used for extensive calculations, L for selected cases.

The substrate is modeled with either 2 or 4 SiC layers, to properly account for the deep deformation possibly induced by larger vacancies (hereafter labeled with an additional 2l and 4l suffix). The cubic poly-type 3C-SiC(0001) is used here, while a discussion of the substrate structure effects will be reported in a forthcoming paper, including also the effects of doping simulated by polarizing and charging the system. Here we report simulations in neutral conditions.

*The vacancy models*

S and L models are combined with vacancies of different sizes from 1 to 13 missing H atoms, of different shapes and locations (see Fig 1b). The single H vacancy could be located in different relative positions with respect to the graphene lattice. Fig 1b reports only the two extreme ones, namely the "hollow" (1Hh) and "top" (1Ht). Clearly, as the number of missing atoms increases, the possible combinations of top-hollow sites increase. Fig 1b reports the combinations here chosen among the most "symmetric". 1H vacancies with all possible relative translations appearing in the S cells were additionally considered as well as several combinations of not subsequent 2H vacancies (see also the SI for a full list of studied cases). Some non-symmetric conformations were also considered for completeness.

*Evaluation of vacancy structure and energy*

The structures were first optimized without vacancies. The relative rotation of graphene with respect to SiC is determined by the chosen model, while the relative translation is initially set in such a way that at least one hollow and one top site exist in the L model case. In the S model, it is not possible to have both exactly top and hollow sites in a single cell, therefore two different models are build including an exactly hollow (Sh) and top site (St), respectively. The details of structural parameters are reported in the SI, with a representation of the optimized structures. The relaxation returns a basically undistorted SiC lattice both for the 2l and 4l models. The sheet rippling is negligible (see the SI, section S.3), indicating that in absence of H coverage vacancies, the interaction between graphene and substrate is very weak, and supporting the definition of "quasi free standing monolayer graphene" for this system. The distances of graphene from H and Si are 2.650±0.005Å and 4.141±0.004Å, respectively, in line with previous DFT calculations[15] and about 0.05 Å underestimated with respect to experimental structural data[14]. Considering that the experimental data are taken at room temperature, while calculations reflect a low temperature situation, this discrepancy is reasonable. The evaluated energy difference between St and Sh models is negligible (~0.4meV in favor of Sh).

Structures with vacancies were obtained extracting H atoms in the proper locations and re-optimizing the geometry. We evaluated the energies of the vacancies $E_v$ according to the following formulas

$$E_v = (E_{opt} + nE_H) - E_{full}$$

$E_{opt}$ and $E_{full}$ being the optimized energies of the system with vacancy and with full H coverage respectively, $E_H$ the energy

of isolated H (and n the number of H missing atoms in the vacancy).

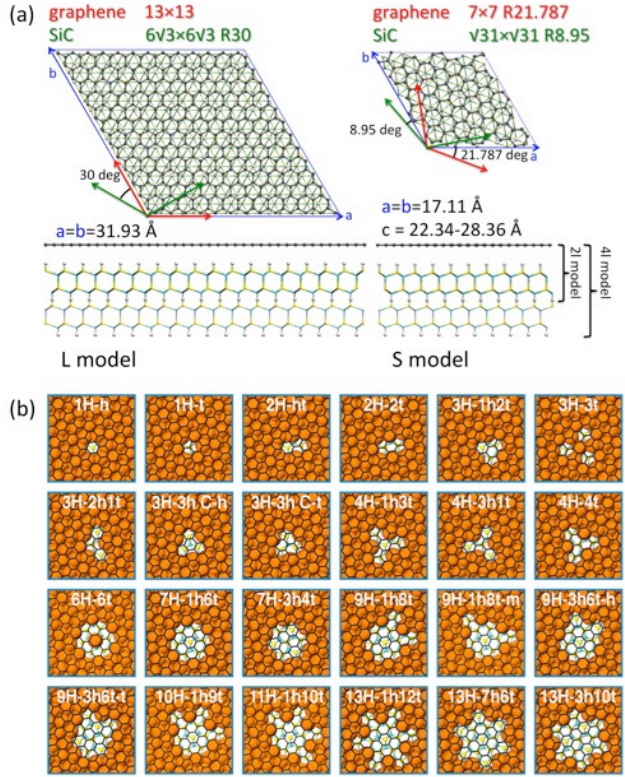

**Fig 1.** (a) Illustration of the model systems. On the left, top and side views of the L-model, including 338 graphene atoms, 864 Si/C substrate atoms (i.e. 432 C and 432 Si for the 4 layers model or 432 – 213 Si and 231 C – for the 2 layers model), and 108 intercalated H atoms and 108 H terminating the bottom side (total 1418, or 986, atoms). On the right the S-model, including 98 graphene atoms, 248 (or 124) substrate Si/C atoms and 31+31 H atoms, (total 408, or 284, atoms). Supercell vectors are represented in blue (length reported). The red and cyan arrows represent the directions of the graphene and SiC unit cell vectors (their rotation with respect to super-cell parameters is reported). (b) Illustration of the type of H vacancies classified according to their size (H atoms are represented in orange as enlarged vdW spheres). Si atoms are in yellow, C substrate atoms in cyan. The labels indicate the number of vacant sites, followed by the number of Si sites located in position top or hollow with respect to graphene (represented in black).

*Calculations Setup*

Calculations are performed with a setup previously tested on this kind of systems[5]. The Rappe-Rabe-Kaxiras-Joannopoulos ultrasoft (RRKJUS) pseudopotentials[17] were used with the Perdew-Burke-Ernzerhof (PBE) exchange-correlation functional[18]. Van der Waals interactions are added within the semi-empirical Grimme D2 scheme[19] (PBE-D2 approach). The plane wave cutoff energy was set to 30 Ry and the density cutoff at 300 Ry (other details of the calculation setup are in the SI, section S.2).

The total density of states (DoS) was evaluated for all cases (local DoS in selected cases). Grids generated according to Monkhorst and Pack scheme[20] were used for integration over the Brillouin Zone: a 10×10×1[21] grid for the S cells and 5x5x1 for the L cell, in order to uniformly sample the corresponding Brillouin zones. Calculations were performed using Quantum ESPRESSO[22] (QE, version 5.3.0).

STM images were simulated at different biases, according to the Tersoff-Hamann theory[23]. QE returns a 3D map of the density of electronic charge integrated within a given energy range delimited by the Fermi level and the given voltage bias. STM images are then generated either by means of "volumes slice" representations of the 3D map, using a horizontal slice plane located just above the graphene sheet, emulating constant height images, or color-coding iso-charge surfaces according to height, which emulates iso-current images. AFM like-images were generated similarly but using the total charge density.

### 2.2 Experimental methods

The theoretical results of this work are compared with STM images of QFMLG samples. These were prepared with a similar procedure as described in Ref. [13]: a 4H-SiC(0001) substrate sample was cleaned by annealing in $H_2$ at 33mbar and 1500 °C for 5 min. A buffer layer was formed by annealing in Ar atmosphere at 800 mbar and 1650 °C for 5 min. The sample was then annealed in $H_2$ at 1013 mbar and at $T_H$= 800 °C for 60min for H intercalation, namely at a lower temperature than in Ref. [13]. The sample was then mounted in an ultra-high vacuum chamber with a base pressure of $1\times10^{-10}$ mbar, and characterized by a RHK Technology STM at room temperature in constant-current mode. At variance with Ref. [13], however, the sample was annealed at 400°C for 14 hours and 600°C for 2 min before the STM observation. This sample is likely to have a higher number of vacant H sites, than the sample studied in Ref. [13], because a lower H intercalation temperature results in less complete H coverage[12], and because of possible further desorption of H during annealing, known to starts at 600°C in vacuum[24],[25].

## 3. Results
### 3.1 Energetic and structural properties
*Single and double H vacancies energies*

Single H vacancies in top (1Ht) and hollow (1Hh) configurations were evaluated in the Sh-2l and St-2l systems, respectively. The vacancy formation energy is +5.169 eV and +5.182 eV for the 1Ht and 1Hh vacancies respectively, i.e. unfavorable, but slightly more favorable (~13meV) for the top-type vacancy. The dependency of vacancy formation energy on the relative translation of the vacant site with respect to graphene was systematically studied by extracting H atoms from each of the single locations. These data show intermediate energy values for sites with intermediate translation between top and hollow (a "map" of the formation energies is reported in the SI). The average value of energies is +5.174 ±0.003eV.

The double H vacancy energies were evaluated for a large number (362) of pairs of 1H vacancies of the Sh-2l model, either adjacent or not, and show a dependence both on the kind of sites (top/hollow) and on the distance between vacant sites, as reported in Fig 2(a) (black dots). On average, the energy is lower for nearer vacancies, and reaches double of the single vacancy energy already for a distance corresponding to the fourth neighbor shell (d~8Å). This indicates a tendency of vacancies to dimerize, with an energy gain of up to ~70meV when vacancies are in the first neighbor shell with respect to the isolated vacancy case. The data corresponding to the first neighbor shell indicate that the most favorable conformation is 2Hth (top-hollow) followed by the 2H2t (two top adjacent sites, 30 meV less stable). All hybrid intermediate configurations are less favored. For larger distances, where the interaction between vacancies is smaller, configurations with prevalence of top character are favored as for the isolated vacancies.

*Energies of larger vacancies*

The reported calculations give indications on the possible structures of larger vacancies: because th and tt adjacent pairs are favored, it is likely that combinations of these will be energetically favored over others. The vacancy types up to 13H built fulfilling this rule are listed in Fig 1b and are those considered in this work. Fig 2b reports $E_v/n_v$ for vacancies of increasing size, $n_v$ being the number of vacant sites, evaluated with S and L models (black and red dots) and with 2l or 4l models (empty and filled dots). Though the general trend as a function of the vacancy size is clear, already for relatively small vacancies (3-4H), 4l models are systematically more stable than 2l models by about 0.1 eV. We impute this to the deformation of the substrate produced around the vacancy, which extends deeper in the substrate Therefore, 4l models should be considered more reliable than 2l ones especially for larger vacancies. The comparison between L (red) and S (black) models, conversely, does not show relevant differences, in this range of vacancy sizes.

The main trend of energy per vacant atom as a function of the vacancy size, however, is clear from Fig 2b: energies decrease up to the size of ~7H, then appear to rise again towards a plateau, the only outlier being the 3H case with separate single H vacancies. This indicates that single vacancies tend to aggregate, which makes the configuration with less but larger vacancies more likely than configuration with many single H vacancies. The differences in energy for vacancies larger than 4H are very small (> 10 meV), which makes hopping and aggregation in larger vacancies possible in appropriate environmental conditions. To complete the view we evaluated the $E_v/n_v$ in the case of complete dehydrogenation (i.e. for the buffer layer), which turns out to be ~4.53 eV. This indicate that for some critical size – presumably that at which the vacancies start to merge within each other – the energy will start again to decrease towards the buffer value.

*Structural properties*

In all cases, the sheet displays an inward bending centered over the vacancy. Overall, the bending increases as the vacancy size increases, although with a large spread as shown in Fig 3a, which reports the maximum vertical displacement Δh of the sheet as a function of vacancy size. We attribute the spread to the variability of the vacancy structures (the correlation of Δh with other possible measurements of the dip – e.g. the local inward curvature – is reported in the SI). Δh can be correlated with AFM measurements, which were used – combined with STM data – to associate features observed in our previous work[13] with vacancies of size 3H or 4H. Fig 3a also reports simulated AFM images of selected cases, generated from the total charge density from DFT calculations. Due to bending, the vacancies are seen as dark spots, whose contrast increases with vacancy depth. In the S model, the vacancy density corresponds to the maximum possible, with vacancies located on a ~6×6 SiC lattice. In the L model with a single vacancy per supercell, conversely, they are located on the sites of a 6√3×6√3 SiC lattice, and the global density is ~1/3 of the previous case (see top right inset in Fig 3a).

For 1H or 2H vacancies, the contrast is very small, practically undetectable. Larger vacancies are generally deeper and more contrasted, although the spread in Δh also causes contrast variations within the same size range. In particular, we observe that the 13H1h-12t ("star") vacancy has a larger contrast in the S model than in the L model (and correspondingly the Δh of the L model is sensibly smaller, as visible by red and black dots comparison in Fig 3a). We attribute this difference to the larger contraction of the sheet in the S cell (0.4%), which favors the inward bending. We finally observe that, because the sheet tends to smoothen the irregularities of the H vacancy boundaries, the AFM images appear basically circular even for triangular or star-shaped vacancies. Therefore AFM is not the ideal tool to detect the shape of the vacancy: it catches only the structural features of the graphene sheet profile, which can be roughly described as an elastic membrane pulled down in the center of the vacancy.

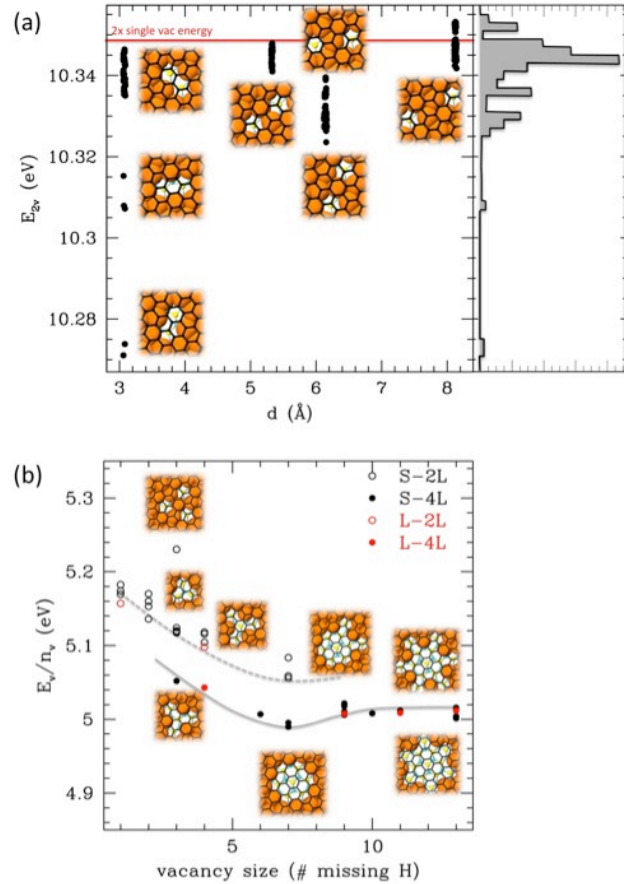

**Fig 2.** (a) 2H vacancy energy as a function of the distance $d$ between vacant sites (black dots). Sample configurations of vacancies are reported. The histogram on the right reports the distribution of energy values. (b) Vacancy formation energy per vacant atom as a function of the vacancy size. Energies are evaluated for different models, and represented with different symbols, as explained in the legend. Selected structures are reported near their corresponding points. Lines are guides to the eye.

Fig 3b reports the minimum Si-graphene distance as a function of the vacancy size. A similar spread as that for Δh is observed, enhanced in this case by the fact that the minimum C-Si distance also depends on the centering of the vacancy (on a top or on a hollow graphene site). In spite of this, there is a visible trend indicating that the distance decreases from a starting value of ~4.2 Å, corresponding to that of the flat sheet without vacancy, to ~3.3 Å, less than the vdW distance be-tween Si and C (3.8 Å). This indicates that the graphene-Si interaction is stronger than a barely physical one, due to the presence of dandling bonds, but weaker than a real chemical bond, whose length would be ~1.9 Å. The Si-gr distance decreases both because the graphene bends and because the Si atom tends to protrude towards it. Therefore the global vacancy energy results from a balance between the bending penalty and the attraction of Si. Data of Fig 2b shows that the latter dominates up to size ~7, then the former starts to compensate due to excessive bending of graphene.

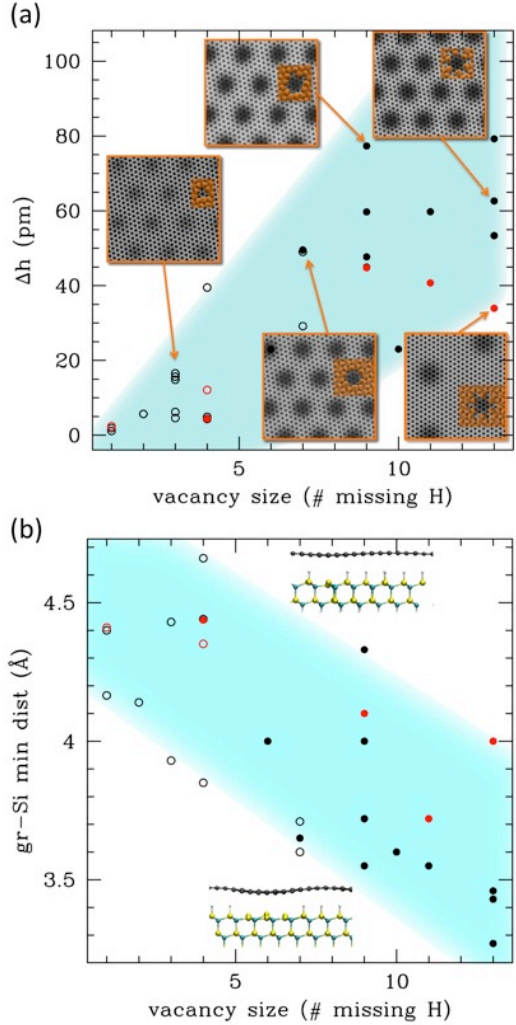

**Fig 3.** (a) Inward bending as a function of vacancy size. The insets report AFM-like images obtained from the total electronic charge of the system (charge levels visualized on a plane located in proximity of the graphene sheet. The same maximum range of grey levels is used for all images). Also reported in each inset is a representation of the H coverage layers showing the vacancy structure. (b) Graphene-Si minimum distance as a function of the vacancy size. Representative structure profiles are reported (for the 3H and 7H, vacancy). Same symbol coding as in Fig 2b.

## 3.2 Electronic properties

### Electronic Density of States

We evaluated the electronic DoS of the system with vacancies, reported in Fig 4a, together with that of single layer isolated graphene for comparison (black solid line). Because of the substrate, the DoS exceeds that of graphene for energies less than -1eV and greater than 1.2 eV with respect to the Dirac point, and the excess DoS is linearly increasing with the number of layers. These energy limits correspond to the top of the valence band and the bottom of the conduction band of SiC. Therefore in the plot the models with 2l or 4l are easily distinguishable, displaying smaller and larger DOS in the SiC bands region. In spite of this, however, the π edges of graphene are still generally visible, which helps aligning the Dirac points of different structures.

In the presence of vacancies, localized states appear, manifesting as peaks near the Dirac point, within the band gap of SiC. They are better visible in the local DoS (Fig 4b) obtained integrating in the region of space strictly including the vacancy, to reduce the contribution of substrate and graphene. Their number, shape, and location depend on the vacancy type. The 1H vacancies display a single peak, located across the Dirac point, corresponding to a half-filled state. As the vacancy size increases, the number of localized states increases, but, due to symmetry induced degeneracy, they group into two or three broader peaks, one still pinned to the Dirac point, the second located in the positive energy region (empty), and a third, when present, located in the negative region (filled). The energy of the empty peak increases with vacancy size, though not very regularly (see also Fig 4c). In addition, for a given vacancy size (same color lines in Fig 4a) the form of the peak is variable, but depending mainly on the shape of the vacancy rather than on its position with respect to graphene lattice. For instance, the three triangular 3H vacancies display practically superimposed peaks in spite of their different rotation and translation with respect to graphene, and different from the elongated and "separated" ones; the star-shaped 13H vacancy has the peak shape different from the other two 13H with different form. Therefore: the location of the "empty state" peak gives an indication on the size of the vacancy, its shape on the vacancy type. This is a generalization of the findings of our previous work[13] where STS measurements revealed peaks at different positive bias, which we associated to 3H and 4H vacancies. The alignment of the Dirac points also allows us to estimate the doping level, which is irregularly dependent on the vacancy type and size, with values in the range 0.07-0.3eV (always n-type).

### Dangling bonds and STM like images

We previously observed[13] that, while AFM-like images always display a negative contrast in correspondence to the vacancies, STM images display a variety of behavior with dark or bright contrast, depending on bias voltage. This is because AFM behavior follows the global structural profile of the sheet, while STM is able to explore selected electronic states, those included in the energy range determined by the bias voltage: positive voltage selects empty states and negative voltage selects filled states. Fig 5 reports in the second column representations of empty (orange) and filled (cyan) states for a series of selected vacancies with missing H≥3 (reported in the first column; the other vacancies are reported in the SI). As can be seen, they appear as states localized between the SiC and the graphene sheet, with lobes protruding upwards (the "dangling bonds"). Lobes of the empty states are generally more protruding, filled ones generally more intruding into the substrate. In addition, the shapes and symmetries of filled and empty states are usually different: e.g. in the 3H reported vacancy, the filled state has two lobes, while the empty one has three; the 7H vacancy empty state has hexagonal symmetry, while the filled one has triangular symmetry; the most evident difference in the larger vacancies is that the empty state is void of charge in the center, while the filled one is not. More complex differences appear in the un-symmetric vacancies.

Fig 5 reports STM-like images in columns 3 to 5. We produced images integrating charge in three different intervals including the localized state peaks in all cases: from the Fermi level to + and –0.5eV and to +1eV. This allows exploring both the structure of the filled and half-filled state pinned to the Fermi level, and the structure of the empty state located at ~1eV. As in the experimental case, the figure shows a variety of contrast types. We can interpret them based on the following considerations:

1. The contrast depends on the balance between the sheet structural inward bending and the protrusion of the electronic states: both bending and protrusion increase as the vacancy size increases, however, protrusion is stronger for the empty states. Therefore positive voltage images are brighter and rich in details characterizing the empty states.

2. Contrast also depends on the iso-current level. Low levels show brighter contrast because they explore low charge level, and therefore the more protruding part of the dangling bond lobes.

3. Images at larger bias are generally brighter, but at those energies a relevant portion of graphene states contribution come again into play. These can be seen at high current, where the π system of graphene is again visible, showing a negative contrast, which follows the graphene bending.

4. The comparison between positive and negative bias images reveals the difference in shape and symmetry of the filled and empty states. We highlight the change from "dimer like" to triangular for the 3H vacancy, the change from triangular to hexagonal symmetry in the 7H vacancy, and the formation of an evident "hole" in the center of the 13H vacancies in the empty state, which give to the star-shaped 13H vacancy the typical "flower" shape (last row).

5. The comparison between images of the same vacancy in S and L models reveal similar shape, but with significant differences. First the different "concentration": the 13H "flower" vacancy nearly "touch" in the S cell. This indicates that this is probably a limiting size for the vacancies, and explains why the "flowers" are rather rare and found only isolated: realizing the maximum concentration with such large vacancies produces a strong rippling.

6. The comparison of same vacancies in the cell shows a smaller inward deformation of the latter, probably due to the larger strain in the L cell, which can cause differences in the images, which are evident for instance comparing 9H1h8t in the two cells. In the large cell, the dark central contrast is practically absent. We attribute this to the different strain level: the S cell favors inward bending.

In summary, the variety of shapes and the contrast dependence on voltage/current levels and on local strain already explains the variety of observed features in STM experiments. In the following section we attempt a more precise assignment based on a direct comparison between simulations and experiment.

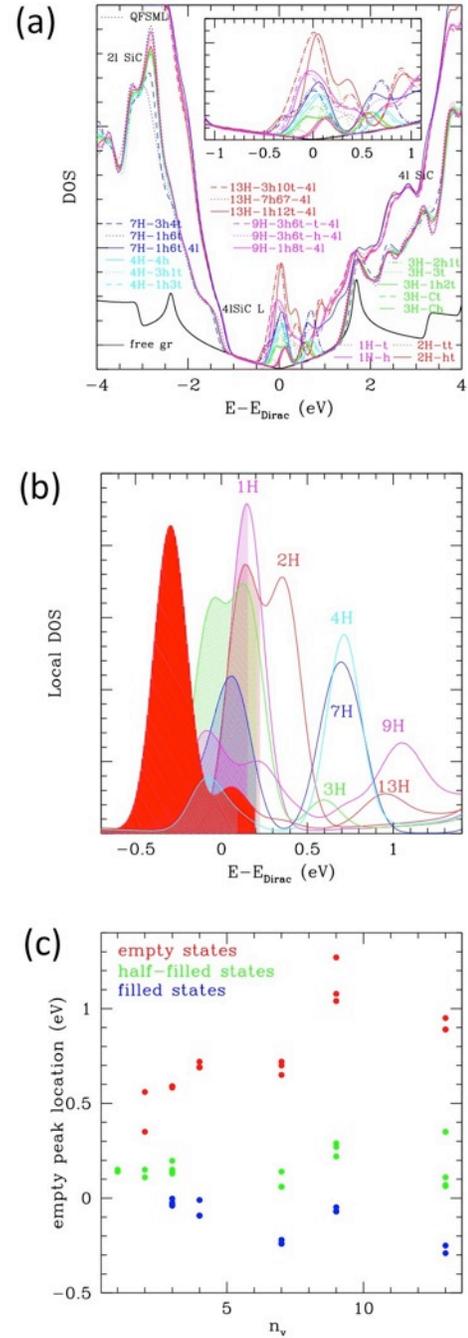

**Fig 4** (a) DOS and (b) local DOS of the system. The DOS of isolated graphene and QFMLG without vacancies are in black (solid and dotted line, respectively), the cases with vacancies are in colors as reported in the plot. When not indicated otherwise, calculations are performed with the S cell with 2l. Only one calculation is with the L cell with 4l (for the 13H-1h12l vacancy), and the corresponding line is labeled with 2lSiC L. DoS are aligned to the Dirac point. (b) Same as (a), but for local DoS, and for selected cases. Color coding as in (a). Normalization of peaks is not uniform, depending on the integration cell used. Some of the peaks are also renormalized for better visualization. Peaks are shaded up to the Fermi energy. (c) Energy location of the peaks as a function of the vacancy size. Empty, filled and half filled states are red, blue and green respectively.

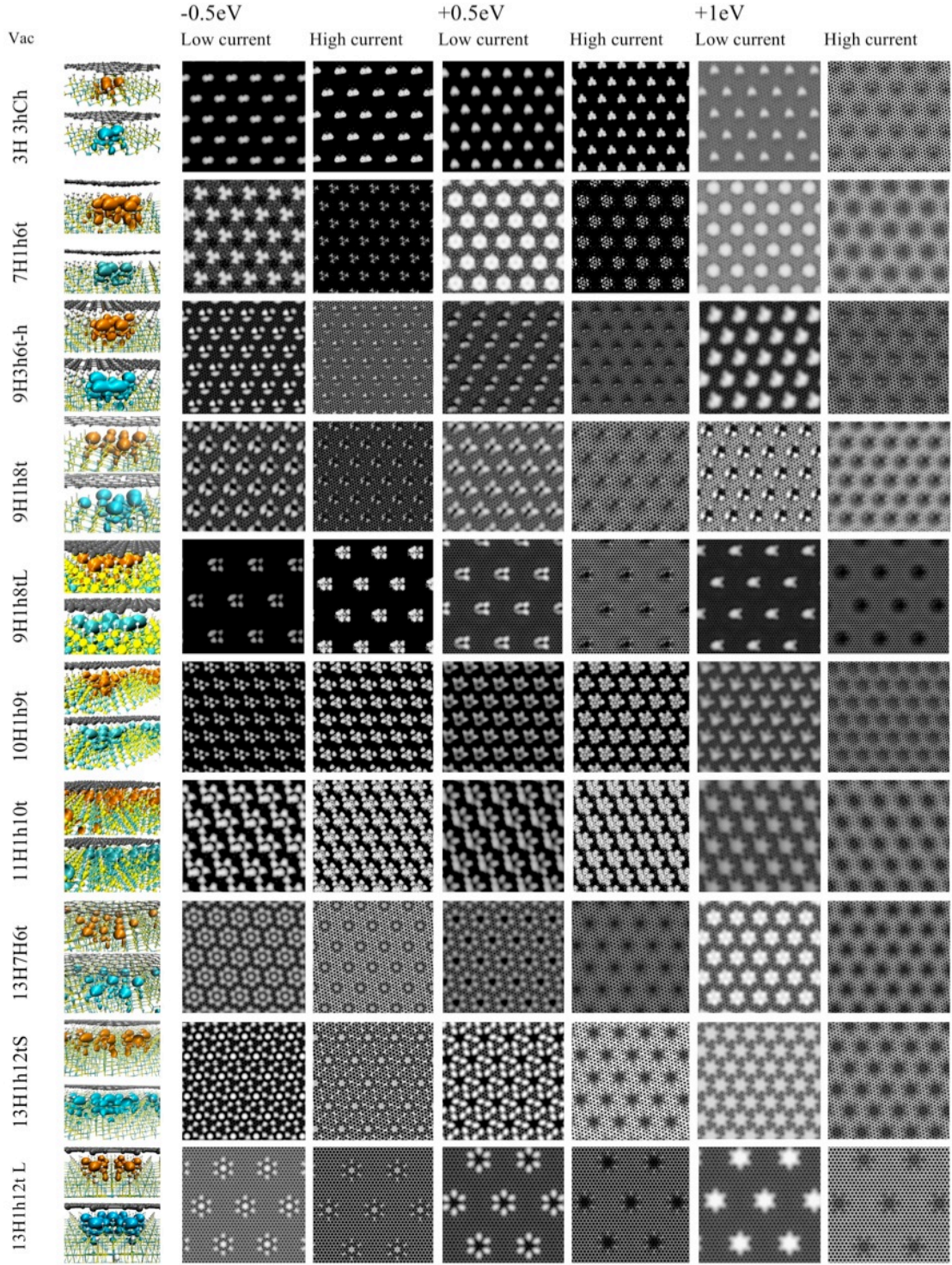

**Fig 5** Localized states and their STM simulated images for selected vacancies of increasing size. The vacancy type is reported in the first column. The second column reports a representation of filled (cyan) and empty (orange) states, obtained integrating the electronic density in the range [0,-0.5] and [0,+0.5] eV with respect to the Fermi level and represented as iso-charge surfaces (iso-value 0.003 in a.u.). Columns 3 to 5 report iso-current images of filled and empty states with integration levels indicated in the header, obtained creating iso-charge images at two different charge levels ($10^{-6}$ a.u. for the low current-like images, left image, and $10^{-4}$ a.u. for the high current-like images, right image). Surfaces are then color-coded according to their z location (bright for protruding). The contrast is adjusted for better visualization.

## 3.3 Comparison with experimental data

Fig 6 reports two STM images from the sample described in section 2.2. In contrast to our previous experiments [12,13], for these measurements the sample was intentionally annealed up to a temperature at which the hydrogen starts to desorb, in order to have lower H coverage. Fig 6 reports experimental STM images in orange (simulated STM images are in gray). We first of all observe that the distance between tightly grouped vacancies in the experimental image is compatible with the maximum density of vacancies realized in model S (see panel (a) and panel (b) in Fig 6, generated with vacancies 7H and 13H-9H respectively). Rotation of the alignment lines of vacancies and distances correspond, and the symmetry is roughly compatible with both L and S model (supercells in white and green superimposed in the left image), with no evident preference for one of the two possible symmetries.

In other regions of the sample, the vacancy concentration is lower. For instance, in the blue circle, vacancies locate on a "hexagon" which roughly corresponds to 2/3 of full occupation (obtainable putting 2 vacancies in L supercell, see panel (c)). Interestingly enough, the flower like vacancy is rather rare, and mostly observed "isolated". Our calculation allows undoubtedly assigning it to the 13H1h12t star-shaped vacancy (panel (d)), due to its easily recognizable shape. Petals point in the armchair directions, apparently with no exception and correspond to the location of the six external vacant H. We infer that it is found isolated because it is quite large and might destructively interact with vacancies of similar size when located on a "maximum occupancy" superlattice (as in panel (b)). "Incomplete" flowers are more frequent, with 2-petals ("Mickey Mouse") or 4-petals ("butterfly"). We assign the "butterfly" to 11H1h10t (panel (e)) and some of the more irregular ones (including the "Mickey") to 9H1h8t (panels f, f') and to 10h1h9t (g, g'). In these vacancies, the external lobes have six equivalent orientations, which in fact seem to appear in different images; in addition, they display large variability in the brightness of the lobes. Overall, we obtain a good correspondence between the features observed in STM and simulation for specific types of vacancies, of size 7H, 9H, 10H, 11H and 13H. The larger vacancy size observed here with respect to our previous work is most probably due to the different hydrogenation procedure generating a lower H coverage. Indeed, our results indicate that this favors the formation of larger vacancies, instead of producing a larger number of smaller ones.

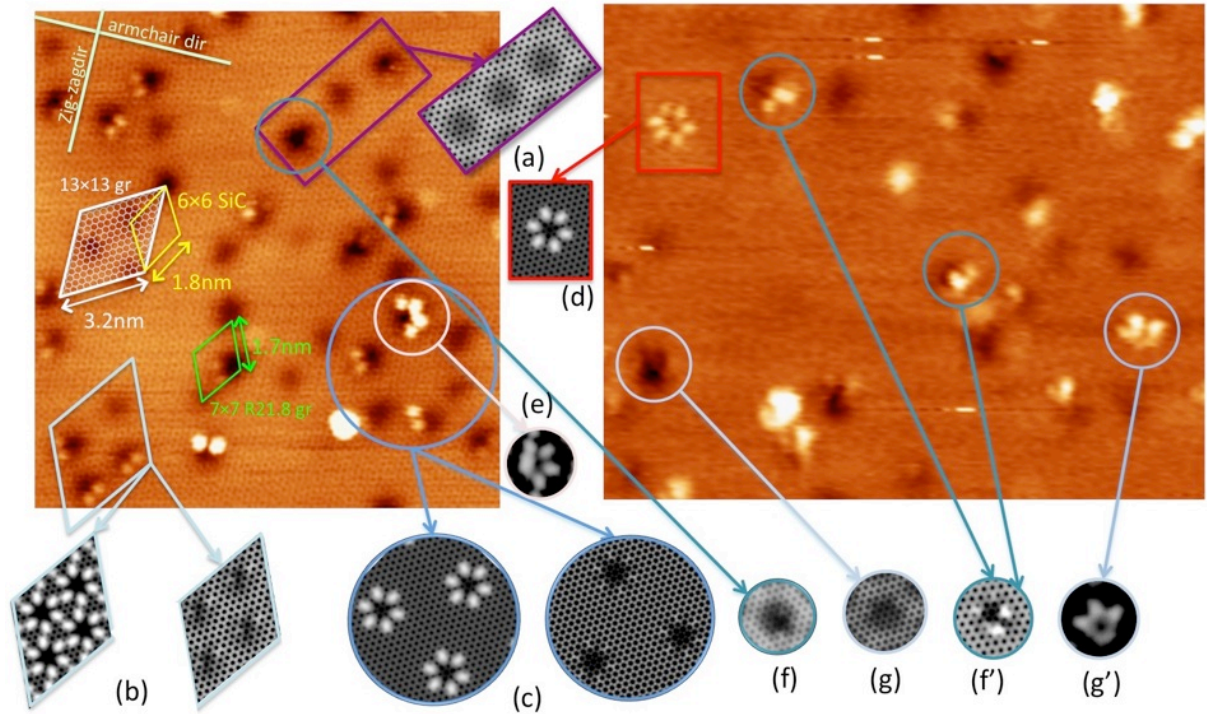

**Fig 6** Comparison between measured and simulated STM images. Measurements are with orange background, taken at bias voltage of 1 V and tunneling current of 1 nA; the sizes of the images are 16.4 nm × 17.6 nm (left) and 20 nm × 16.5 nm (right). STM simulated data are in gray. Panels report portions of the images of Fig. 5, of the following cases: (a) 7H1h6t; (b) 13H1h12t and 9H1h8t (c) and (d) 13H1h12t (e) 11H1h10t (f, f') 9H1h8t (g, g') 10l1h9t. Images are aligned based on zigzag direction clearly visible both in theoretical and experimental data. In the left image the model superlattices are reported as indicated.

## 4. Summary and Conclusions

In our previous report[13] we inferred that, considering the lateral size, depth, and contrast dependence on voltage, observed in the vacancies of the sample examined therein, the most probable vacancy size was of 3 or 4 missing H. Here, a systematic DFT study of vacancies up to 13H size and comparison with new STM data, allowed assigning features ob-served in STM images to larger vacancies, in the range 7H-13H. These can be recognized based on their typical shape and different contrast. We infer that the formation of larger vacancies is due to two circumstances: first the lower H coverage induced by the environmental conditions, second the tendency of single H vacant site to aggregate, here clearly demonstrated by the behavior of the vacancy formation energy as a function of the vacancy size. Therefore when H coverage is low, the system prefers to form larger vacancies than to increase the number of smaller ones.

It is also observed that vacancies tend to locate on a ~6x6 SiC lattice, though not everywhere the maximum density is realized. The size of vacancies (i.e. number of missing H atoms per vacancy) and in some case their precise shape, can be inferred from measured STM images. Conversely, we found little dependence of the simulated STM images and of their energies on the relative translation between the vacancy and graphene lattice. At first sight this might seem incompatible with the observation that the vacancies tend to locate on a regular lattice. In fact, it is not: the hydrogenation process proceeds from the buffer layer, which has its own quasi-6x6 symmetry, descending from the $6\sqrt{3}\times6\sqrt{3}$. Therefore it is natural that the vacancies, resulting from the last hydrogenated sites, also follow this symmetry, in spite of the relative translation of graphene.

From a methodological point of view, we showed that not only the "standard" $6\sqrt{3}\times6\sqrt{3}$ supercell model is compatible with a correct representation of this system, but also the 1/3 smaller $\sqrt{31}\times\sqrt{31}$ supercell, which allows much faster calculations. In this, the relative rotation between SiC and graphene is only <1 deg different from the "standard" one, and the graphene little contracted. The comparison between theory and experiment shows that the two model systems give very similar results. We can therefore conclude that most experimental features discussed in this paper are well described already by the small model, at a much lower computational expense compared to the large model. We remark however, that besides the slightly different symmetry, the two models generate a different strain of the graphene sheet. This results in a slightly different behavior of the STM simulated contrast, especially in the less symmetric vacancies. A similar variability is observed indeed in the STM data, which might indicate that small local variation of the strain of the graphene sheet might occur in the real system. Indeed, a number of interesting questions remain to be investigated, such as the full hydrogenation/dehydrogenation energy profile in different environmental conditions and the role of the buffer and of the local strain in the hydrogenation process. These studies are ongoing and will be the matter of a forthcoming work about the hydrogenation process.

The electronic properties of this system turn out very interesting. The vacancies display localized states corresponding to dangling bonds. With the exception of the (undetectable) smallest ones, these group generally in two (or three, for the largest ones) energy peaks, one pinned to the Fermi level, one in the positive energy region, and therefore empty. The empty states are those more easily detected by STM and generating characteristic images. The energy location of these states increases with the size of the vacancy. In addition the energy separation between the empty states and the filled ones is of the order of ~2-3 eV and tunable with vacancy size, still located in between the two π edges of graphene.

The energy tunability and the possibility of accurately locate these states are pre-requisites for their exploitation in devices for coherent nano- opto-electronics

### Supporting Information

A PDF file including details of calculation setup, numerical data in tabular form and additional results.
A zipped archive including the optimized atomic structures in .xyz format

The Supporting Information is available free of charge on the ACS Publications website.

### Corresponding Author

* E-mail: valentina.tozzini@nano.cnr.it; Phone: +39 050 509433

### Current address

† Kwansei Gakuin University, 2-1 Gakuen, Sanda, Hyogo 669-1337, Japan,

### Funding Sources

## ACKNOWLEDGMENT

We thank Prof. Paolo Giannozzi, Dr. Camilla Coletti, and Dr. Vittorio Pellegrini for useful discussions. We gratefully acknowledge financial support by EU-H2020, Graphene-Core1 (agreement No 696656 ), MCSA (agreement No 657070), by CINECA awards IsB11_flexogra (2015), IsC36_ElMaGRe (2015), IsC44_QFSGvac (2016), IsC44_ReIMCGr (2016) and PRACE "Tier0" award Pra13_ 2016143310 (2016). We aknowlegde CINECA staff for technical support. SH acknowledges travel support from COST Action MP1103 "Nanostructured materials for solid-state hydrogen storage".

# Revealing the nature of defects in quasi free standing monolayer graphene on SiC(0001) by means of Density Functional Theory

Tommaso Cavallucci,[1] Yuya Murata,[1] Makoto Takamura,[2] Hiroki Hibino,[2,†] Stefan Heun,[1] Valentina Tozzini[1,*]

[1] NEST, Istituto Nanoscienze-CNR and Scuola Normale Superiore, Piazza San Silvestro 12, 56127 Pisa, Italy
[2] NTT Basic Research Laboratories, 3-1 Morinosato Wakamiya, Atsugi, Kanagawa 243-0198, Japan

# Supporting Information

List of supporting information files

1. this pdf file
2. a zipped archive `Structures.zip`, whose expansion deliver the following subdirectories

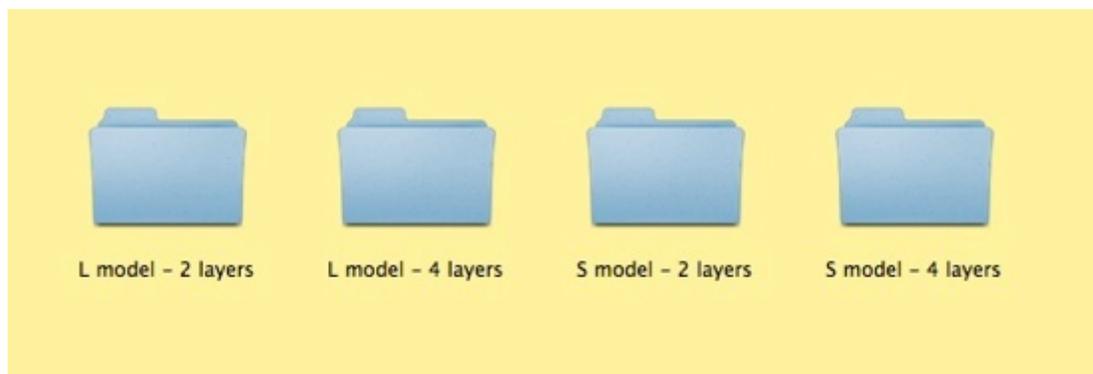

Each folder includes the unit cell of the optimized structures in .xyz format of the models indicated in the folder name and a file with the supercell parameters (`CellParameters`) to build the superlattices.

## S.1. List of performed calculations

A summary of performed calculations in the different model systems is reported in table S.1. The 1H structural calculations were performed over all the 31 sites of the two S cells (row 1H 30 sites), while the calculations for the 2H couples are 362, chosen with vacancies at different distances (see main text).

|  | L model | | S model | | | |
|---|---|---|---|---|---|---|
|  | | | Translation "top" (St) | | Translation "hollow" (Sh) | |
|  | 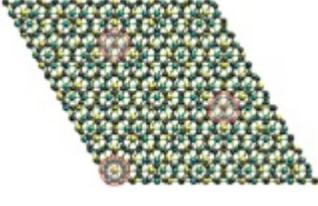 | | 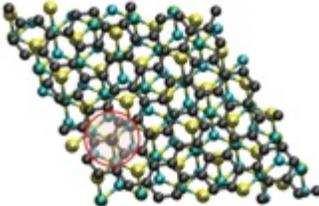 | | 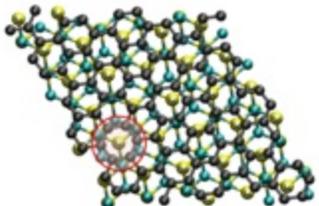 | |
|  | 4 SiC layers | 2 SiC layers | 4 SiC layers | 2 SiC layers | 4 SiC layers | 2 SiC layers |
| No vac | Str opt | Str opt | Str opt | Str opt | Str opt, DoS | Str opt, DoS |
| 1H-t |  |  |  | Str opt, DoS, STM |  |  |
| 1H-h | Str opt |  |  |  |  | Str opt, DoS, STM |
| 1H 30 sites |  |  |  | Str opt |  | Str opt |
| 2H-ht |  |  |  |  |  | Str opt, DoS, STM |
| 2H-tt |  |  |  | Str opt, DoS, STM |  |  |
| 2H mix |  |  |  | Str opt |  | Str opt |
| 3H-3t |  |  |  |  |  | Str opt, DoS, STM |
| 3H-1h2t |  |  |  |  | Str opt, DoS, STM | Str opt, DoS, STM |
| 3H-2h1t |  |  |  | Str opt, DoS, STM |  |  |
| 3H-3h C |  |  |  | Str opt, DoS, STM |  | Str opt, DoS, STM |
| 4H-1h3t |  |  |  |  |  | Str opt, DoS, STM |
| 4H-3h1t |  |  |  | Str opt, DoS, STM |  |  |
| 4H-4t | Str opt | Str opt, DoS |  | Str opt, DoS, STM |  |  |
| 6H-6t |  |  |  |  | Str opt, STM |  |
| 7H-1h6t |  |  |  |  | Str opt, DoS, STM | Str opt, DoS, STM |
| 7H-3h4t |  |  | Str opt, DoS, STM | Str opt, DoS, STM |  |  |
| 9H-1h8t | Str opt, STM |  |  |  | Str opt, DoS, STM |  |
| 9H-1h8t-m |  |  |  |  | Str opt, STM |  |
| 9H-3h6t |  |  | Str opt, DoS, STM |  | Str opt, DoS, STM |  |
| 10H-1h9t |  |  |  |  | Str opt, STM |  |
| 11H-1h10t | Str opt, STM |  |  |  | Str opt, STM |  |
| 13H-1h12t | Str opt, DoS, STM |  |  |  | Str opt, DoS, STM |  |
| 13H-7h6t |  |  |  |  | Str opt, DoS, STM |  |
| 13H-3h10t |  |  | Str opt, DoS, STM |  |  |  |

**Table S.1.** Tabular summary of the performed calculations. St = S model with translation of type "top" Sh = S model with translation of type "top". The calculations performed in each case are reported: Str opt= structural optimization; DoS= Density of states, STM=STM-like calculations (see next sections). In the header a top view of the L model and of the two S models are reported. Sites with t and h translations are highlighted with circles.

## S.2. Details of calculations
### S.2.1 Local DoS

In addition to the DOS, also the local DOS (LDOS) was calculated, using different space boxes centered on a vacancy. Quantum ESPRESSO requires that the box sides are parallel to the cell axes, so each box is a prism with rhomboidal base (see Figure 3). Two different box basis in the xy plane were chosen: the first one, designed for the 1H vacancy, has a length of 2.38 Å and an height of 2.28 Å (small box, see fig S1), while the second, designed for the 7H, has a length of 9.51 Å and an height of 8.56 Å (large box). The z dimension was varied depending on the number of atomic layers that we wanted to include in the box.

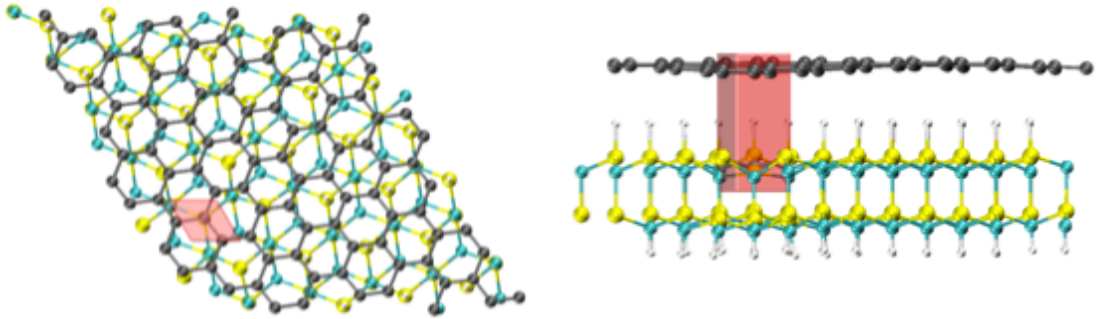

**Fig S.1.** LDOS calculation box shape in red (small box is reported). Each side of the box must be parallel to a cell vector.

## S.2.2 Measure of curvature

The local curvature is measured through the out of plane pseudo dihedral defined as shown in Fig S.2 a. For all cases, this angle is measured locally on the vacancy ($\phi_l$) and its root mean squared deviation over all structure ($\sqrt{\phi_l^2}$), which is a measure of globar rippling. A third measure of the deformation due to the vacancy is the total vertical deformation $\Delta h$ (i.e. maximum-minimum vertical displacement of C atoms of the graphene sheet). A correlation plot of the three measures of deformation is reported in Fig S2.

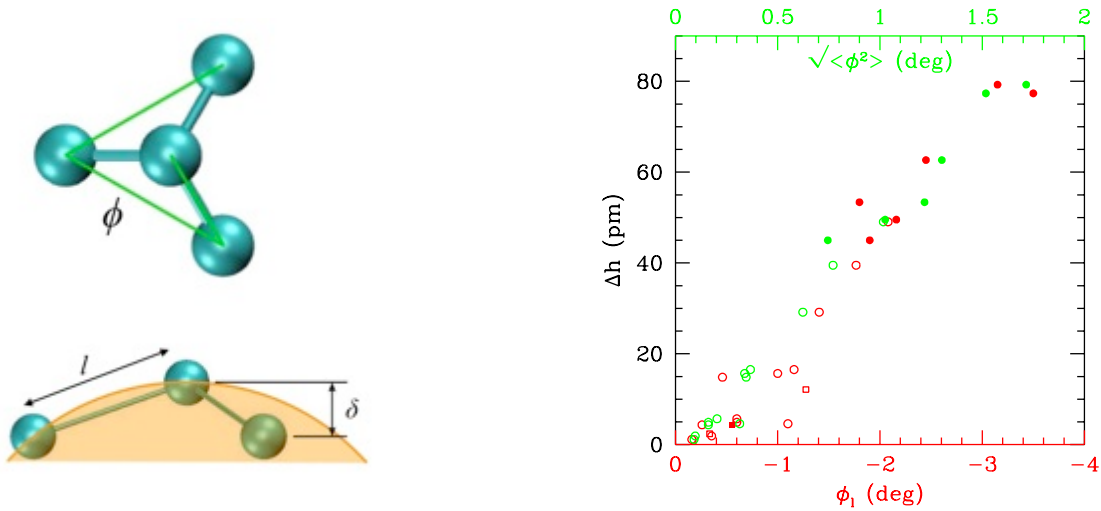

(a)          (b)

**Fig S.2.** (a) Illustration of the out of plane $\phi$ dihedral around for a given C atom, as the improper dihedral between the sites connected by the green lines and definition of the distance $\delta$, the bond length parameter $l$. A sketch of the tangent sphere is reported in orange. (b) correlation plot of the local (red) and global (red) dihedral angle with the vertical displacement. Empty and filled dots correspond to 4l and 2l models respectively

## S.2.3 Other details

Gaussian smearing is set at 0.01 Ryd in all calculations. The convergence threshold for self-consistency was set to $10^{-8}$. The BFGS quasi-Newton algorithm was used for structural optimizations[1], with standard convergence criteria, i.e. $10^{-3}$ a.u. for the forces and $10^{-4}$ a.u. for the energy. Structural optimization were performed with Γ only sampling of the Brillouin zone.

## S.3. Additional results

### S.3.1 Optimized structures with full coverage

Side views of the St-2l, Sh-2l and Sh-4l models are reported in Fig S.2 (top views are in Table S.1).

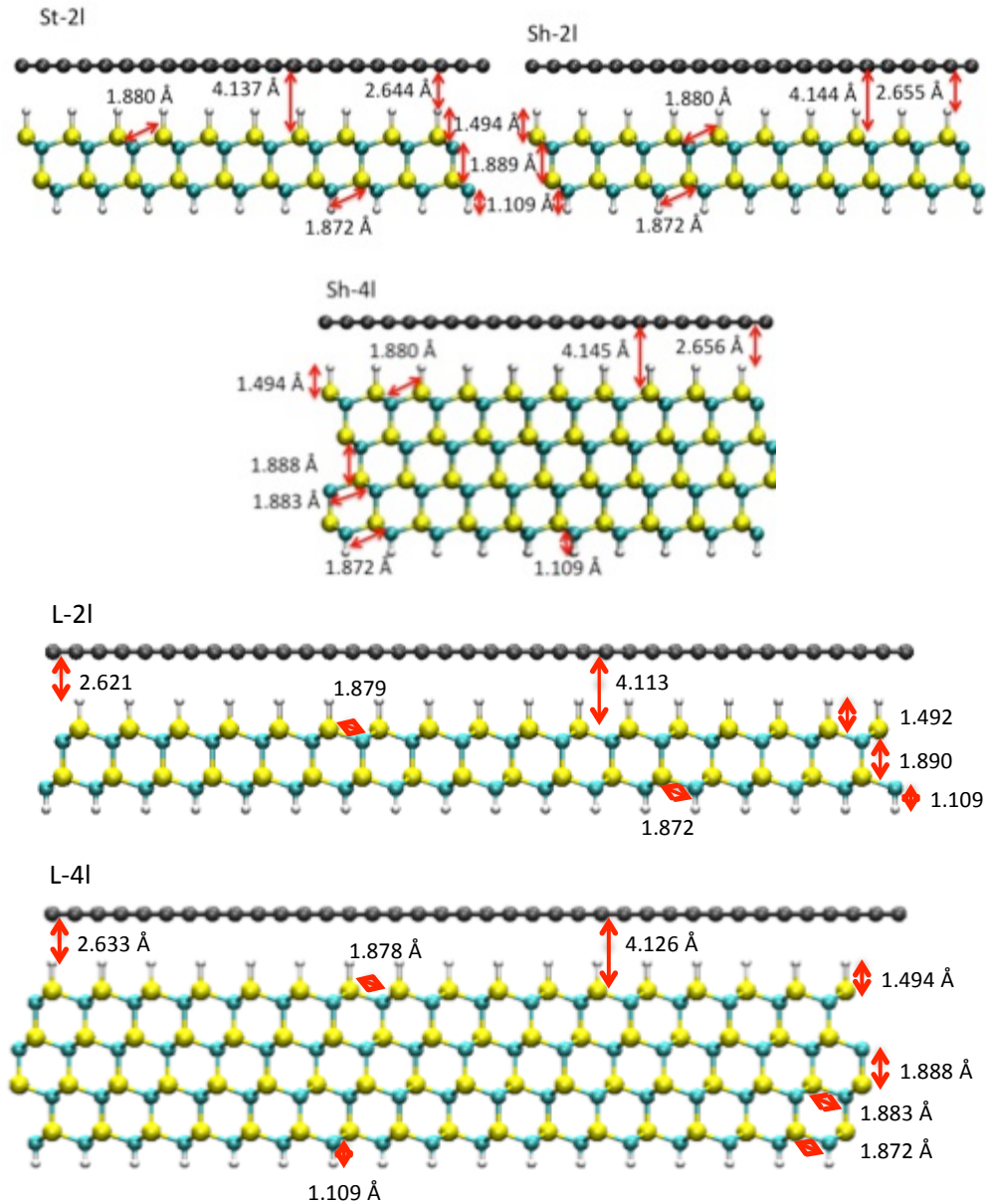

**Fig S.2**. From top to bottom: Side views of the optimized St-2l Sh-2l, Sh-4l, L-2l and L-4l with bond lengths reported. The average sheet bending is negligible ($\sqrt{\langle \varphi^2 \rangle} \sim 0.079$ deg).

### S.3.2 Energies and some structural parameters

The formula for the evaluation of the vacancy energy is already reported in the main text. We remark however, that we can take advantage of the calculation procedure to separate the energy vacancy in a mechanical and a "chemical part". In fact, vacancies are generated from the QFSMLG by extracting single H atoms from the saturating layer, and subsequently optimizing. Therefore we also have the energy of the structure before optimization $E^*$, which is "purely chemical" as opposed to the final one where there is a gain for relaxation. Therefore one can write

$$E_v = (E_{opt} + nE_h) - E_{full} = [(E^* + nE_h) - E_{full}] + (E_{opt} - E^*) = E_{chem} + E_{mech}$$

The numerical values of these energies (and of their value per unit atom) are in Table S.2. Interestingly enough, $E_{chem}/n$ are much less variable than $E_v/n$ as a function of the vacancy size assuming the values ~5.18 and ~5.12 for basically all the 2l and 4l models, respectively. Therefore, the variation of the "mechanical" energy per vacancies follows the same trend of the $E_v/n$ displaying a decrease and a minimum at n~7 (see fig. S.3). Table S2 also reports other structural parameters, which are plotted in Fig 3 in the main text.

Figure S.4 reports a color map of the single vacancy formation energy for the S models (31 values for each): bright values correspond to lower energy values, dark to higher energy values. i.e. the vacancy can appear more easily on the bright sites. We remark that these energies are with respect to "atomized" hydrogen, i.e. the removed atom is in atomic form. This explains the high values of $E_{chem}/n$ and $E_v/n$. There

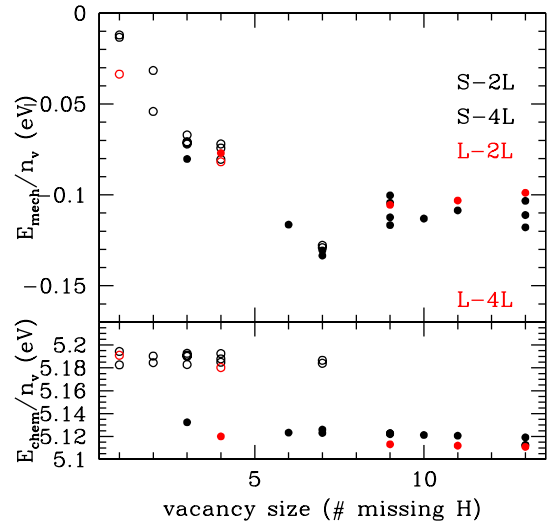

**Fig S.3**. Chemical (bottom) and mechanical (top) energy per missing atom. The legend for symbols is within the plot.

are other possible reference level: e.g. using as reference the molecular hydrogen, all energies would decrease of a value corresponding to half the binding energy of $H_2$; alternatively, one could consider as reference energy that of the buffer layer, which would imply a subtracting offset of ~4.53 eV, corresponding to the energy per vacant site in case of full dehydrogenation. All of these alternative representations are rigid offsets: the relative energies per vacant site remain unchanged, as far as their behavior as a function of the vacancy size.

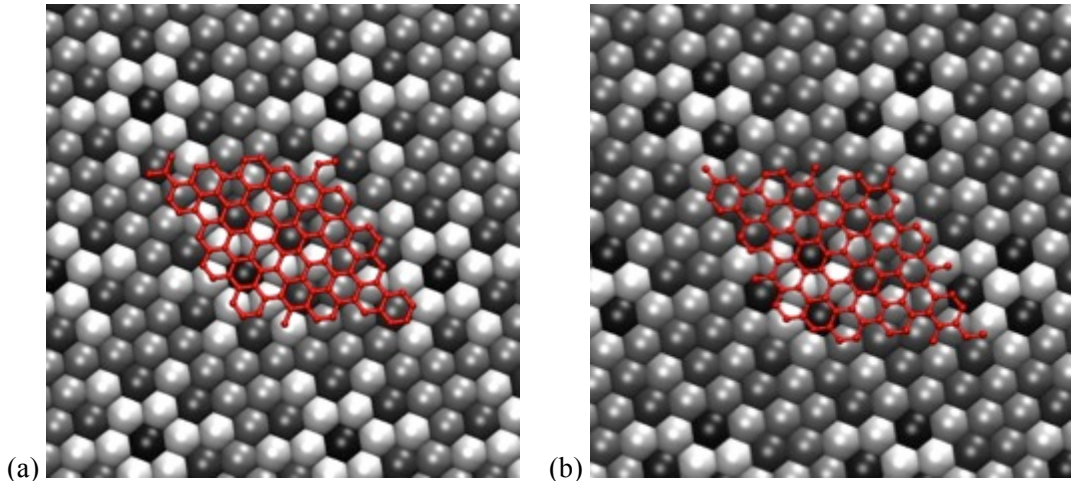

(a) (b)
**Fig 4.** (a) Mapping of the energy formation of 1H vacancies (model Sh-l2) onto the H layer (represented as enlarged vdW spheres). Color coding: black=+5.182 eV, white =+5.169 eV, shades of gray: intermediate values in linear scale. The graphene lattice (a single supercell) is superimposed to show the relative translation with the underneath H layer. (b) Same but with model St-2l.

| Mod/vac | $E_{vac}$ (eV) | $E_{chem}$ (eV) | $E_{str}$ (eV) | Si-$C_{vac}$ (Å) | $\langle\sqrt{\varphi^2}\rangle$ (deg) | $\Delta h$ (Å) |
|---|---|---|---|---|---|---|
| St-2l | 5.1694 | 5.1827 | -0.0133 | 4.165 | 0.0969 | 0.0190 |
| 1Ht | 5.1728 | | -0.00985 | 4.037 | 0.2877 | 0.1337 |
| Sh-2l 1Hh | 5.1823 | 5.1943 | -0.0120 | 4.40 | 0.0909 | 0.0114 |
| | 5.1861 | | -0.00820 | 4.34 | 0.1841 | 0.0845 |
| Avg (over all sites) | 5.174±0.003 | | | | | |
| L-2l 1Hh | 5.1574 | 5.1910 | -0.03355 | 4.41 | 0.08386 | 0.024 |
| St-2l 2Htt | 10.305912=2×5.153 | 10.36904=2×5.1845 | -0.06313 | 4.14 | 0.2036 | 0.0568 |
| avg | 10.32 =2×5.16 | | | | | |
| Sh-2l 2Hht | 10.2724=2×5.1362 | 10.3806=2×5.1903 | -0.10822 | 4.12 | 0.2134 | |
| Sh-2l 3H1h2t | 15.3548=3×5.1183 | 15.5709=3×5.1903 | -0.21614 | 3.93 | 0.3143 | 0.046 |
| Sh-4l 3H 1h2t | 15.1565=3×5.0522 | 15.3973=3×5.1324 | -0.24081 | 3.97 | 0.2757 | |
| St-2l 3H2h1t | 15.3720=3×5.1240 | 15.5732=3×5.1911 | -0.20111 | 4.19 | 0.20001 | |
| St-3Ht-2l | 15.6921=3×5.2307 | 15.5487=3×5.1829 | 0.14336 | 4.07 | 0.346 | 0.1484 |
| Sx 3H C centered top | 15.3521=3×5.1174 | 15.5658=3×5.1886 | -0.21367 | 3.97 | 0.3386 | 0.1567 |
| Sx 3H C centered hollow | 15.3600=3×5.1200 | 15.5724=3×5.1908 | -0.21239 | 3.99 | 0.3668 | |
| Sh-2l 4H1h3t | 20.4632=4×5.1158 | 20.7512=4×5.1878 | -0.28799 | 4.06 | 0.1611 | |
| St-2l 4H3h1t | 20.4731=4×5.1183 | 20.7699=4×5.1925 | -0.29683 | 4.25 | 0.1726 | |
| St-2l 4Ht | 20.4179=4×5.1045 | 20.7398=4×5.1850 | -0.32194 | 3.93 | 0.7706 | 0.3950 |
| L-2l 4Ht | 20.3939=4×5.0985 | 20.7210=4×5.1803 | -0.32714 | | | |
| L-4l 4Ht | 20.1725=4×5.0431 | 20.4806=4×5.1201 | -0.30803 | | | |
| Sh-4l 6Ht | 30.0424=6×5.0071 | 30.7406=6×5.1234 | -0.69820 | | | |
| Sh-2l 7H1h6t | 35.2106=7×5.0301 | | | 4.25 | 1.0051 | 0.4975 |
| Sh-2l 7H1h6t (frozen layer) | 35.3846=7×5.0549 *35.5858=7×5.0837* | 36.2880=7×5.1840 | -0.90343 | 3.81 | 1.0166 | 0.4904 |
| Sh-4l 7H1h6t | 34.9278=7×4.9897 | 35.8617=7×5.1231 | -0.93381 | 3.83 | 1.0256 | 0.4953 |
| St2l 7H3h4t | 35.3040=7×5.0434 | | | 4.1 | 0.8508 | 0.4436 |
| St2l 7H3h4t (frozen layer) | 35.4121=7×5.0589 *35.3932=7×5.0562* | 36.3066=7×5.1867 | -0.89447 | 4.04 | 0.6229 | 0.2916 |
| St4l 7H3h4t | 34.9683=7×4.9955 | 35.8822=7×5.1260 | -0.91391 | 3.70 | 0.4118 | |
| Sh-4l 9H1h8t | 45.1954=9×5.0217 | 46.0980=9×5.1220 | -0.90262 | 3.65 | 1.1871 | |
| L-4l 9H1h8t | 45.0695=9×5.0077 | 46.0192=9×5.1132 | -0.94969 | | | |
| Sh-4l 9H1h8t meta | 45.0549=9×5.0061 | 46.1045=9×5.1223 | -1.04967 | | | |
| Sh-4l 9H3h6t | 45.1646=9×5.0183 | 46.1037=9×5.1126 | -0.93912 | 3.72 | 0.7450 | |
| St-4l 9H3h6t | 45.0959=9×5.0107 | 46.1070=9×5.1230 | -1.01106 | 3.51 | 1.5187 | |
| Sh-4l 10H1h9t | 50.0828=10×5.0083 | 51.2133=10×5.1213 | -1.13055 | | | |
| Sh-4l 11H1h10t | 55.1338=11×5.0122 | 56.3276=11×5.1207 | -1.19379 | | 1.2001 | |
| L-4l 11H1h10t | 55.0987=11×5.0090 | 56.2330=11×5.1121 | -1.13432 | | | |
| Sh-4l 13H1h12t | 65.2068=13×5.0159 | 66.5497=13×5.1192 | -1.34284 | | 1.3027 | |
| L-4l 13H1h12t | 65.1573=13×5.0121 | 66.4425=13×5.1110 | -1.28525 | | 0.3464 | |
| Sh-4l-13H7h6t | 65.0541=13×5.0042 | 66.5865=13×5.1220 | -1.53241 | 3.74 | 1.7151 | 0.7926 |
| St-4l-13H3h10t | 65.0206=13×5.0016 | 66.4651=13×5.1127 | -1.44448 | 3.76 | 1.3027 | 0.6266 |

**Table S.2** Structural parameters of the optimized structures with vacancies. The distance Si-$C_{vac}$ is evaluated in the center of the vacancy for the "top" case, or in the central hexagon in the "hollow" case. $\varphi_{vac}$ is the out of plane dihedral evaluated using the central vertex in the "top" case or the 6 central vertices in the "hollow" cases. $\Delta h$ is the maximum vertical displacement of the graphene sheet, a measure of its deformation.

### S.3.3 STM – like images

STM-like images for vacancies not reported in the main text are reported in Table S.3. Besides the iso-current images generated with the same protocol as that described in Fig 5 in the main text, here samples of STM-like images at fixed height are also reported for comparison in selected cases, generated placing a plane near the graphene surface and color-coding it according to the local density value. This was done only for the empty state (at +1eV).

| Vac | | Iso-height | Iso-current (top = low current, bottom = high current) | | | |
|---|---|---|---|---|---|---|
| | | +1eV | -1eV | -0.5eV | +0.5eV | +1eV |
| 1Hh | | | | | | |
| 1Ht | | | | | | |
| 2Hht | | | | | | |
| 2Htt | | | | | | |
| 3H3t | 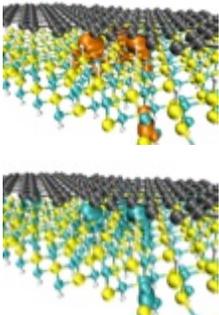 | 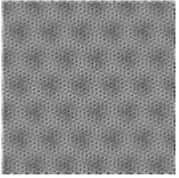 | 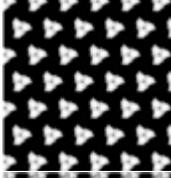 | 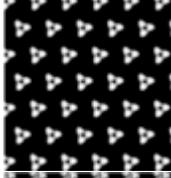 | 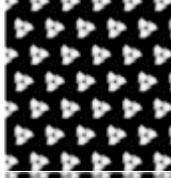 | 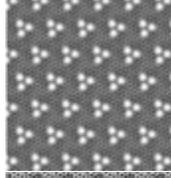 |
| 3H1h2t | 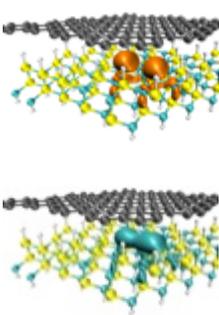 | 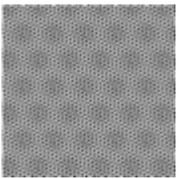 | 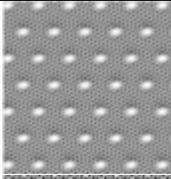 | 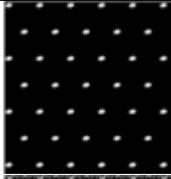 | 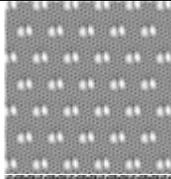 | 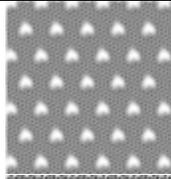 |
| 3H2h1t | 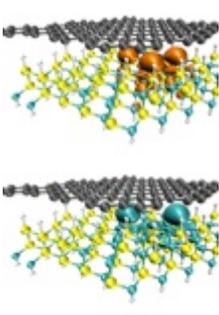 | 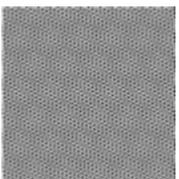 | 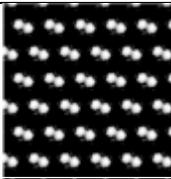 | 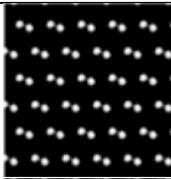 | 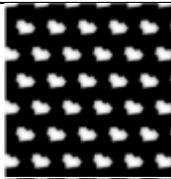 | 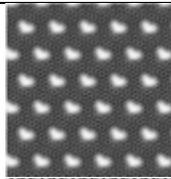 |
| 3H3hCt | 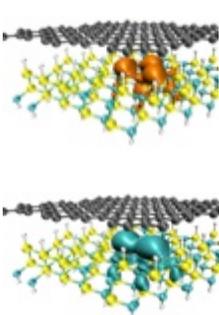 | 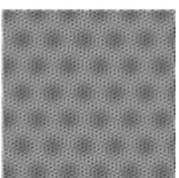 | 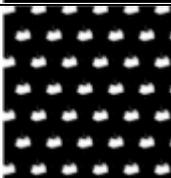 | 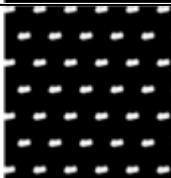 | 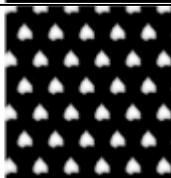 | 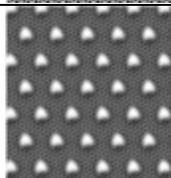 |
| 4H1h3t | 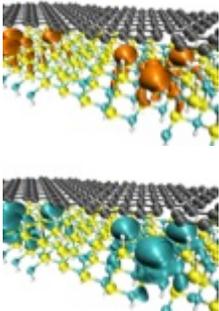 | 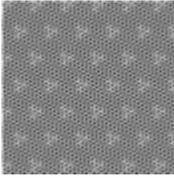 | 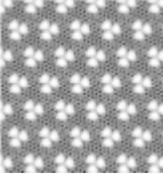 | 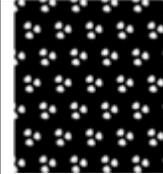 | 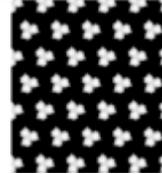 | 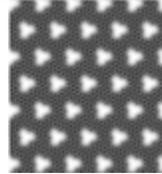 |
| 4H3h1t | 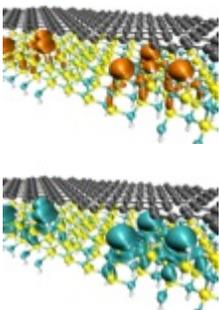 | 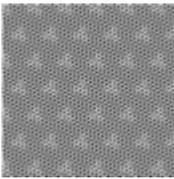 | 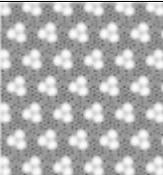 | 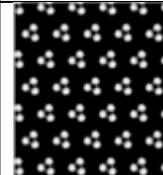 | 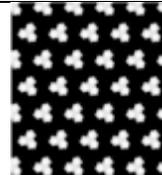 | 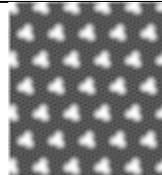 |
| 4H4t | 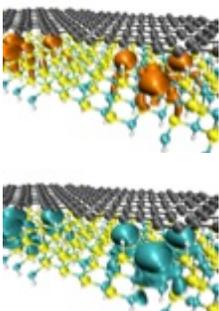 | 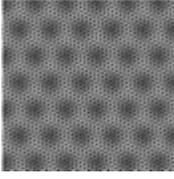 | 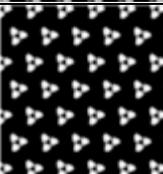 | 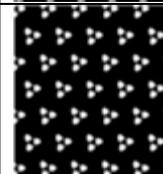 | 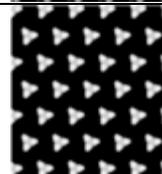 | 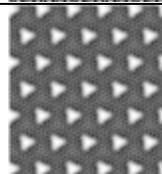 |
| 6H6t | 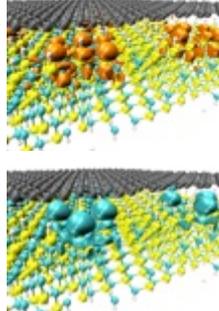 | 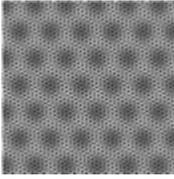 | 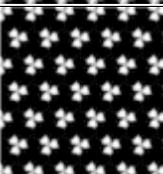 | 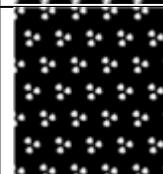 | 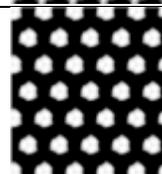 | 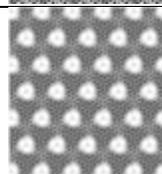 |

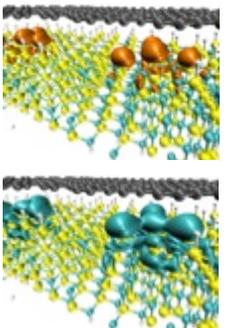

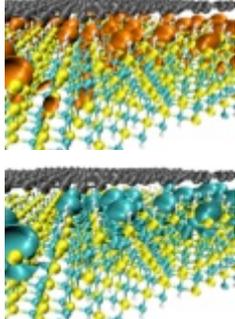

**Table S.3** Localized states and their STM simulated images for vacancies not reported in the main text. The vacancy type is reported in the first column. The second column reports a representation of filled (cyan) and empty (orange) states, obtained integrating the electronic density in the range [0,-0.5] and [0,+0.5] eV with respect to the Fermi level and represented as iso-charge surfaces (iso-value 0.003 in a.u.). Third column reports selected iso-height images generated at +1V and using a plane at <1 Å from graphene sheet. Columns 4 to 6 report iso-current images of filled and empty states with integration levels indicated in the header, obtained creating iso-charge images at two different charge levels ($10^{-6}$ a.u. for the low current-like images (left image) and $10^{-4}$ a.u. for the high current-like images (right image)). Surfaces are then color-coded according to their z location (bright for protruding). The contrast is adjusted for better visualization.